\documentclass[10pt,a4paper]{article}
\usepackage{amsmath,amsfonts,amssymb,amscd}
\catcode`\@=11
\@addtoreset{equation}{section}

\catcode`\@=12
\def\interior{\,\hbox{\vrule depth0pt height.6pt width4pt \vrule depth0pt height8pt}\;\,}
\def\de#1/de#2{\frac{\partial {#1}}{\partial {#2}}}

\def\E{{\cal E}}
\def\C{{\cal C}}
\def\J{{\cal J} ({\cal E}\times {\cal C})}
\begin{document}
\title{\textbf{A square-torsion modification of Einstein-Cartan theory}}
\author{Stefano Vignolo$^{1}$\footnote{E-mail: vignolo@diptem.unige.it}\,, Luca Fabbri$^{1,2}$\footnote{E-mail: fabbri@diptem.unige.it}
\, and Cosimo Stornaiolo$^{3}$\footnote{E-mail: cosmo@na.infn.it}\\
\footnotesize{$^{1}$DIME Sez. Metodi e Modelli Matematici, Universit\`{a} di Genova}\\
\footnotesize{Piazzale Kennedy Pad. D, 16129 Genova, Italy}\\
\footnotesize{$^{2}$INFN \& Dipartimento di Fisica, Universit\`{a} di Bologna}\\
\footnotesize{Via Irnerio 46, 40126 Bologna, Italy}\\
\footnotesize{$^{3}$INFN Sez. di Napoli, Compl. Univ. Monte S. Angelo Ed. N}\\
\footnotesize{Via Cinthia, I- 80126 Napoli, Italy}}
\date{}
\maketitle
\begin{abstract}
\noindent In the present paper we consider a theory of gravity in which not only curvature but also torsion is explicitly present in the Lagrangian, both with their own coupling constant. In particular, we discuss the couplings to Dirac fields and spin fluids: in the case of Dirac fields, we discuss how in our approach, the Dirac self-interactions depend on the coupling constant as a parameter that may even make these non-linearities manifest at subatomic scales, showing different applications according to the value of the parameter we have assigned; in the case of spin fluids, we discuss FLRW cosmological models arising from the proposed theory.
\end{abstract}
\section{Introduction}
The issue about the torsional completion of gravity has already covered several decades, and still it is not fully resolved: although torsion is a natural part of the most complete metric-compatible connection, nevertheless there are also reasons against torsion that are admittedly justified. In this paper, we are going to bring new insights in this topic.

As we have just mentioned, once the covariant derivative is written in the most general way, it is defined upon a connection that in general is not symmetric in the two lower indices, and therefore torsion does not generally vanish; although the implementation of the principle of equivalence and causality may seem to impose the symmetry of the connection, and thus restricting torsion to be equal to zero \cite{w,m-t-w}, nevertheless a deeper analysis has shown that even in the case in which these principles hold then torsion is only restricted to be completely antisymmetric \cite{m-l,xy,so,Fabbri:2006xq,Fabbri:2009se}: the inclusion of torsion beside the curvature makes it possible to introduce a torsion-spin coupling beside the curvature-energy coupling, fully realizing the geometry-matter coupling prescription. It is actually quite intriguing that if torsion is completely antisymmetric this complete coupling prescription could only be realized by the one fermion field that we have ever been able to observe at the moment \cite{Fabbri:2008rq,Fabbri:2009yc}; and if we now focus on this single fermion field, namely the Dirac spinor, the torsional effects manifest themselves as self-interactions, capable of giving a dynamical explanation of the exclusion principle \cite{Fabbri:2010rw}.

However, there are also problems that seem to push against the acceptation of torsion: the most important of which is the fact that the torsional effects influence the dynamics only at the Planck scale and not much beyond, making torsion negligible in almost every situation; this is due to the fact that torsion has a coupling constant that is the gravitational constant. The fact that torsion and curvature share the coupling constant might have been thought to have something to do with the fact that both fields are geometrical fields and therefore with the same strength; on the other hand however, despite torsion and curvature come in fact from a common geometrical background, nevertheless they are independent fields, and consequently endowing them with an identical strength is not justifiable \emph{a priori}, and they should in general have different coupling constants. In the present paper, we shall propound what we believe to be one of the simplest theories of gravity where torsion is present with its own coupling constant, showing that the presence of torsion with a properly tuned coupling gives consequences already at subatomic scales.

This theory of gravity is simply based on the idea that, as for a complete description of the underlying geometrical properties not only curvature but also torsion is to be considered, then in order to study the dynamical properties the Lagrangian should not only be written in terms of the curvature but also torsional terms: this extension is natural and in literature it has already been put forward \cite{Baekler:2011jt,Baekler:2006de,Tresguerres:1995un,Tresguerres:1995js, McCrea:1992wa}; in those papers, when all results are accounted, one may see that there are several terms that need be added to the Lagrangian, but if one assumes the least-order derivative hypothesis, then there can only be terms that in the curvature are linear and in torsion are quadratic, so that if one additionally wants to neglect parity-violating terms then one is left with the usual torsionless Lagrangian plus three torsion-squared contributions \cite{Fabbri:2012yg}. Here however we would like to consider only one of these contributions, since we are not interested in developing the most mathematically comprehensive of the models but its physical consequences; thus instead of dwelling in all mathematical details of the most general theory, we shall focus on a specific model so to have a clearer look at the modifications that such a generalization will bring into the phenomenological aspects. As we will see, the presence of this additional quadratic-torsion term will have the effect of producing a shift in the value of the constant with which torsion couples to matter fields, so that the torsional coupling constant is not necessarily the Newton gravitational constant, and its value still to be determined may be large indeed, and in particular one may employ this modification to see that torsional effects may be present in particle physics and amplified in cosmology.

In particular, for particle physics like in \cite{Fabbri:2012yg} we are here going to study the seminal case of Dirac fields, while for the cosmological applications differently from \cite{Fabbri:2012yg} where ELKO have been consider here we are going to employ the spin fluid; in addition we will employ a different formalism \cite{VC,CV,CVB1,CVB2,VCB} to write the structure of the theory.

The layout of the paper is as follows: in section \ref{2} we recall the main features of the $\cal J$-bundles geometry, in which we develop our physical theory; in section \ref{3} we formulate the square-torsion modification of the original Einstein-Cartan theory, showing that in this way curvature and torsion appear in the dynamical field equations each with its own coupling constant; in section \ref{4} we apply this model to two of the most relevant matter fields with spin, the Dirac spinor and the spin fluid: for the Dirac field we discuss how, by choosing a suitable value for the coupling constant of torsion, it is possible to show that, in the case of two leptons, the Dirac equation has self-interactions that can be written in the form of weak forces, while in the case of two neutrinos, those self-interactions can be used to model neutrino oscillation; in the case of the spin fluid, we study cosmological applications, showing how the already known cosmological effects of torsion (absence of initial singularity, initial accelerated expansion, horizon problem) may be affected by the choice of the torsional coupling constant.
\section{The Geometrical Framework}
\label{2}
To begin with, we shall recall the notation and convention we will employ in the following of the paper: readers that are familiar with the formalism of the $\cal J$-bundles may skip this sections and pass directly to the following one; those who are not, may also refer to \cite{VC,CV,CVB1,CVB2,VCB} for further details.

Let us consider a $4$-dimensional manifold with a metric tensor $g$ of signature $\eta=(1,3)=(1,-1,-1,-1)$. Let $\E$ be the co-frame bundle of $M$ and $P\to M$ be a principal fiber bundle over $M$. We denote by $\C :=J_1(P)/SO(1,3)$ the space of principal connections over $P$, referring $\E$ and $\C$ to local coordinates $x^i,e^\mu_i$ ($i,\mu =1,\dots,4$) and $x^i,\omega_i^{\;\;\mu\nu}$ ($\mu < \nu$).

The configuration space of our theory is the fibered product $\E\times_{M}\C$ ($\E\times\C$ for short) over $M$. The dynamical fields are (local) sections of $\E\times\C$, namely pairs formed by a tetrad field $e(x)=e^\mu_i(x)\,dx^i$ and a spin connection $1$-form $\omega(x)= \omega_i^{\;\;\mu\nu}(x)\,dx^i$, automatically metric-compatible with the metric $g(x)=\eta_{\mu\nu}\,e^\mu(x)\otimes e^\nu(x)$. 

We consider the first $\cal J$-bundle ${\cal J}(\E\times\C)$ associated with the fibration $\E\times\C\to M$; it is a fiber bundle built similarly to an ordinary jet-bundle, but the first order contact between sections is considered with respect to exterior differential.

The starting point is be the first jet-bundle $J_1(\E\times\C)$ associated with the fibration $\E\times\C\to M$, referred to local jet-coordinates $x^i,e^\mu_i,\omega_i^{\;\;\mu\nu},e^\mu_{ij},\omega_{ij}^{\;\;\;\mu\nu}$. Then, we introduce on $J_1(\E\times\C)$ the following equivalence relation: let $z=(x^i,e^\mu_i,\omega_i^{\;\;\mu\nu},e^\mu_{ij},\omega_{ij}^{\;\;\;\mu\nu})$ and $\hat{z}=(x^i,\hat{e}^\mu_i,\hat{\omega}_i^{\;\;\mu\nu},\hat{e}^\mu_{ij},\hat{\omega}_{ij}^{\;\;\;\mu\nu})$ be two elements of $J_1(\E\times\C)$, having the same projection $x$ over $M$; denoting by $(e^\mu(x),\omega^{\mu\nu}(x))$ and $(\hat{e}^\mu(x),\hat{\omega}^{\mu\nu}(x))$ two different sections of the bundle $\E\times\C\to M$, respectively chosen among the representatives of the equivalence classes $z$ and $\hat z$, we say that $z$ is equivalent to $\hat z$ if and only if
\begin{subequations}\label{0.0}
\begin{equation}
e^\mu(x) = \hat e^\mu(x), \qquad \omega^{\mu\nu}(x)=\hat{\omega}^{\mu\nu}(x)
\end{equation}
and 
\begin{equation} 
de^\mu(x)= d\hat{e}^\mu(x), \qquad D\omega^{\mu\nu}(x)=D\hat{\omega}^{\mu\nu}(x)
\end{equation}
\end{subequations}
where $D$ denotes the covariant differential induced by the connection. In local coordinates, it is easily seen that $z \sim \hat{z}$ if and only if the following identities hold 
\begin{subequations}
\label{0.00}
\begin{equation}
e^\mu_i=\hat{e}^\mu_i, \qquad \omega_i^{\;\;\mu\nu}=\hat{\omega}_i^{\;\;\mu\nu}
\end{equation}
\begin{equation}
(e^\mu_{ij}- e^\mu_{ji})=(\hat{e}^\mu_{ij}- \hat{e}^\mu_{ji}), \qquad (\omega_{ij}^{\;\;\;\mu\nu}-\omega_{ji}^{\;\;\;\mu\nu})= (\hat{\omega}_{ij}^{\;\;\;\mu\nu}-\hat{\omega}_{ji}^{\;\;\;\mu\nu}).
\end{equation}
\end{subequations}
The quotient space $J_1(\E\times\C)/\sim$ is denoted by ${\cal J}(\E\times\C)$ and the corresponding canonical projection is indicated by $\rho: J_1(\E\times\C)\to {\cal J}(\E\times\C)$. A system of local fibered coordinates on the bundle ${\cal J}(\E\times\C)$ is provided by the set of functions $x^i,e^\mu_i,\omega_i^{\;\;\mu\nu},E^\mu_{ij}:=\frac{1}{2}\left(e^\mu_{ij}- e^\mu_{ji}\right),\Omega_{ij}^{\;\;\;\mu\nu}:=\frac{1}{2}\left(\omega_{ij}^{\;\;\;\mu\nu}-\omega_{ji}^{\;\;\;\mu\nu}\right)$ $(i<j)$. The quotient projection $\rho$ endows the bundle ${\cal J}(\E\times\C)$ with most of the standard features of jet-bundles geometry ($\cal J$-extension of sections, contact forms, $\cal J$-prolongation of morphisms and vector fields), which are needed to implement variational calculus on ${\cal J}(\E\times\C)$. Referring the reader to \cite{VC,CV,CVB1,CVB2} for a detailed discussion on $\cal J$-bundles geometry, the relevant fact we need to recall here is that the components of the torsion and curvature tensors can be chosen as fiber $\cal J$-coordinates on ${\cal J}(\E\times\C)$. In fact, the definitions of torsion and curvature components
\begin{subequations}\label{0.000}
\begin{equation}
T^\mu_{ij}=2E^\mu_{ji} + \omega^{\;\;\mu}_{i\;\;\;\lambda}e^\lambda_j - \omega^{\;\;\mu}_{j\;\;\;\lambda}e^\lambda_i
\end{equation}
\begin{equation}
R_{ij}^{\;\;\;\;\mu\nu} = 2\Omega_{ji}^{\;\;\;\mu\nu} + \omega^{\;\;\mu}_{i\;\;\;\lambda}\omega_j^{\;\;\lambda\nu} - \omega^{\;\;\mu}_{j\;\;\;\lambda}\omega_i^{\;\;\lambda\nu}
\end{equation}
\end{subequations}
can be regarded as fiber coordinate transformations on ${\cal J}(\E\times\C)$, allowing to refer the bundle ${\cal J}(\E\times\C)$ to local coordinates $x^i,e^\mu_i,\omega_i^{\;\;\mu\nu},T^\mu_{ij}$ $(i<j)$, $R_{ij}^{\;\;\;\;\mu\nu}$ $(i<j, \mu <\nu)$. In such coordinates, local sections $\gamma :M\to{\cal J}(\E\times\C)$ are expressed as
\begin{equation}\label{0.1}
\gamma : x \to (x^i,e^\mu_i(x),\omega_i^{\;\;\mu\nu}(x),T^\mu_{ij}(x),R_{ij}^{\;\;\;\;\mu\nu}(x))
\end{equation}
In particular, a section $\gamma$ is said holonomic if it is the $\cal J$-extension $\gamma ={\cal J}\sigma$ of a section $\sigma:M\to \E\times\C$. In local coordinates, a section is holonomic if it satisfies the relations \cite{VC}
\begin{subequations}\label{0.2}
\begin{equation}
T^\mu_{ij}(x) = \de{e^\mu_j}(x)/de{x^i} - \de{e^\mu_i}(x)/de{x^j} + \omega^{\;\;\mu}_{i\;\;\;\lambda}(x)e^\lambda_j(x) - \omega^{\;\;\mu}_{j\;\;\;\lambda}(x)e^\lambda_i(x)
\end{equation}
\begin{equation}
R_{ij}^{\;\;\;\;\mu\nu}(x) = \de{\omega_{j}^{\;\;\mu\nu}(x)}/de{x^i} - \de{\omega_{i}^{\;\;\mu\nu}(x)}/de{x^j} + \omega^{\;\;\mu}_{i\;\;\;\lambda}(x)\omega_j^{\;\;\lambda\nu}(x) - \omega^{\;\;\mu}_{j\;\;\;\lambda}(x)\omega_i^{\;\;\lambda\nu}(x)
\end{equation}
\end{subequations}
namely if the quantities $T^\mu_{ij} (x) $ and $R_{ij}^{\;\;\;\;\mu\nu} (x) $ are precisely the components of the torsion and curvature tensors associated with the tetrad $e^\mu_i (x) $ and the connection $\omega_i^{\;\;\mu\nu} (x) $, in turn representing the section $\sigma $.

We also recall that the bundle ${\cal J} (\E\times\C) $ is endowed with a suitable contact bundle, locally spanned by the following $2$-forms
\begin{subequations}\label{0.3}
\begin{equation}\label{0.3a}
\theta^\mu = de^\mu_i \wedge dx^i + E^\mu_{ij}\,dx^i \wedge dx^j
\end{equation}
\begin{equation}\label{0.3b}
\theta^{\mu\nu} = d\omega_i^{\;\;\mu\nu}\wedge dx^i + \Omega_{ij}^{\;\;\;\mu\nu}\,dx^i \wedge dx^j
\end{equation}
\end{subequations}
It is easily seen that a section $\gamma :M\to{\cal J} (\E\times\C) $ is holonomic if and only if it satisfies the condition $\gamma^*(\theta^\mu)=\gamma^*(\theta^{\mu\nu})=0 $ $\forall \mu,\nu =1,\ldots,4 $. Moreover, in the local coordinates $x,e,\omega,T,R $ the $2$-forms \eqref{0.3} can be expressed as
\begin{equation}\label{0.4}
\theta^\mu = \tau^\mu - T^\mu \qquad{\rm and} \qquad \theta^{\mu\nu}= \rho^{\mu\nu} - R^{\mu\nu}
\end{equation}
being $\tau^\mu = de^\mu_i \wedge dx^i + \omega_{j\;\;\;\nu}^{\;\;\mu}e^\nu_i\,dx^j \wedge dx^i\/$, $T^\mu = \frac{1}{2}T^\mu_{ij}\,dx^i \wedge dx^j\/$, $\rho^{\mu\nu}= d\omega_i^{\;\;\mu\nu}\wedge dx^i + \frac{1}{2}\left(\omega_{j\;\;\;\lambda}^{\;\;\mu}\omega_i^{\;\;\lambda\nu} - \omega_{j\;\;\;\lambda}^{\;\;\nu}\omega_i^{\;\;\lambda\mu}\right)\,dx^j \wedge dx^i\/$ and $R^{\mu\nu}=\frac{1}{2}R_{ij}^{\;\;\;\;\mu\nu}\,dx^i\wedge dx^j\/$.
 
A Lagrangian on $\J $ is any horizontal $4$-form, locally expressed as
\begin{equation}\label{1.0}
L={\cal L} (x^i,e^\mu_i,\omega_i^{\;\;\mu\nu},T^\mu_{ij},R_{ij}^{\;\;\;\;\mu\nu})\,ds
\end{equation}
We can associate with any given Lagrangian a corresponding Poincar\'e-Cartan $4$-form, having local expression (see \cite{VC})
\begin{equation}\label{1.1}
\Theta = {\cal L}\, ds - \frac{1}{2}\de{\cal L}/de{T_{hk}^\alpha}\,\theta^\alpha\wedge ds_{hk}
- \frac{1}{4}\de{\cal L}/de{R_{hk}^{\;\;\;\;\alpha\beta}}\,\theta^{\alpha\beta}\wedge ds_{hk}
\end{equation}
where $ds_{hk}:=\de /de{x^h}\interior\de /de{x^k}\interior ds $. 

Any Lagrangian \eqref{1.0} gives rise to an associated a variational principle of the kind 
\begin{equation}\label{1.2bis}
{\cal A} (\sigma)=\int{\cal J}\sigma^* (\Theta)=\int{\cal J}\sigma^* ({\cal L}\,ds)
\end{equation}
where $\sigma:M\to \E\times\C $ denotes an arbitrary section and ${\cal J}\sigma:M\to \J $ its $\cal J$-extension satisfying eqs. \eqref{0.2}. Referring again the reader to \cite{VC} for details and comments, we just recall that the corresponding Euler-Lagrange equations can be expressed as
\begin{equation}\label{1.4}
{\cal J}\sigma^*\left({\cal J} (X)\interior d\Theta\right)=0
\end{equation}
for all $\cal J$-prolongable vector fields X on $\E\times\C $. Due to the arbitrariness of $X $ and the holonomy of the $\cal J$-extension ${\cal J}\sigma $, from the the requirement \eqref{1.4} we get two sets of final field equations
\begin{subequations}\label{1.7}
\begin{equation}\label{1.7a}
{\cal J}\sigma^*\left(\de{\cal L}/de{e^\mu_q} + \de{\cal L}/de{T_{kq}^\alpha}\omega_{k\;\;\;\mu}^{\;\;\alpha} \right) -\de /de{x^k}\left({\cal J}\sigma^*\left(\de{\cal L}/de{T_{kq}^\mu}\right)\right) =0
\end{equation}
and 
\begin{equation}\label{1.7b}
\begin{split}
{\cal J}\sigma^*\left( \de{\cal L}/de{\omega_q^{\;\;\mu\nu}} - \de{\cal L}/de{T_{kq}^\mu}e^\sigma_k\eta_{\sigma\nu} + \de{\cal L}/de{T_{kq}^\nu}e^\sigma_k\eta_{\sigma\mu} + \de{\cal L}/de{R_{kq}^{\;\;\;\;\alpha\nu}}\omega_{k\;\;\;\mu}^{\;\;\alpha} + \de{\cal L}/de{R_{kq}^{\;\;\;\;\mu\alpha}}\omega_{k\;\;\;\nu}^{\;\;\alpha}\right)\\
-\de /de{x^k}\left({\cal J}\sigma^*\left(\de{\cal L}/de{R_{kq}^{\;\;\;\;\mu\nu}} \right)\right)
=0
\end{split}
\end{equation}
\end{subequations}
\section{The Gravitational Square-Torsion Lagrangian}
\label{3}
The geometrical setting outlined above does not only have the advantage of allowing a geometric meaning to general gauge theories, but also that of highlighting that, as both tetrad and spin connection are independent and they must be treated equally, similarly both torsion and curvature tensors have to be considered on an equal level of importance; this is certainly true during the variation of the Lagrangian that leads to the field equations, and it should therefore not surprise that such a complementarity has to be found in the construction of the Lagrangian itself. This is however not the case in general: for instance, the Einstein-Cartan theory, that is the simplest that can be derived within this formalism, comes from a Lagrangian of the form ${\cal L}_G=\sqrt{|g|}R$ which clearly favours curvature tensor over torsion, despite the fact that both curvature and torsion are taken as dynamical variables in deriving the field equations. So the Einstein-Cartan theory seems too simple.

Inasmuch as both curvature and torsion are taken as independent degrees of freedom, a more general treatment allowing torsion not only implicitly within the curvature but also explicitly on its own appears to be more elegant: an example of theories in which torsion must be considered explicitly and not only implicitly through the curvature so that the Lagrangian is constructed on an intertwined mix between torsion and curvature tensors are the conformal theories of gravity \cite{Fabbri:2011vk}. However, these theories have higher-order derivative Lagrangians, and consequently their torsion can propagate in regions in which the effects of torsion are expected to be vanishingly small \cite{k-r-t}; even in the case in which torsion does vanish so that the field equations of Weyl gravity are recovered \cite{Fabbri:2011gt}, there are nevertheless solutions that do not reduce to those of the known Einstein limit \cite{Fabbri:2008vp, Fabbri:2011zz} and so it is not surprising that discrepancies with observations arise \cite{Flanagan:2006ra}. This is not a peculiarity of conformal gravity but of all higher-order theories of gravitation, and a wise choice might consequently be to look for theories that have least-order derivative Lagrangians so to have chances to recover the proper limit. 

The resulting situation is thus a balance between the issues of mathematical coherence and experimental consistency, the first pushing toward a generalization in which torsion is included beside curvature while the second pulling to remain within the observational constraints, a tension resolved by a gravitational Lagrangian of the least-order derivative in curvature and torsion. A gravitational Lagrangian of this kind, linear in curvature and quadratic in torsion is
\begin{equation}\label{3.1}
{\cal L}_G = e\/R + e\/k\/T 
\end{equation}
where $e=\det(e^\mu_i)\/$, $R=R_{ij}^{\;\;\;\;\mu\nu}e^i_{\mu}e^j_{\nu}\/$ and $T=T^{\;\;\;\sigma}_{pq}T_{\;\;\;\sigma}^{pq}$. In the Lagrangian \eqref{3.1}, the fundamental constant for $R$ is the Newton constant normalized to the unity while the fundamental constant for $T$ is $k$ yet to be determined. This gravitational Lagrangian is supplemented by the matter Lagrangian ${\cal L}_M$ in the full Lagrangian ${\cal L}_G-{\cal L}_M$ as usual. 

Taking the identities $\de e/de{e^\mu_i}=ee^i_\mu\/$ and $\de{e^j_\nu}/de{e^\mu_i}=-e^i_{\nu}e^j_{\mu}\/$
into account, we have 
\begin{subequations}
\begin{equation}\label{1.8a}
\de{{\cal L}_G}/de{e^\mu_i}=ee^i_{\mu}\/R - 2e\/R_{\mu\sigma}^{\;\;\;\;\lambda\sigma}e^{i}_\lambda + ek\left(T_{pq}^{\;\;\;\sigma}T^{pq}_{\;\;\;\sigma}e^i_\mu + 4T^{i\;\;\;\sigma}_{\;\;p}T^p_{\;\;k\sigma}e^k_\mu\right)
\end{equation}
\begin{equation}\label{1.8b}
\de{{\cal L}_G}/de{R_{ki}^{\;\;\;\;\mu\nu}}=2e\/\left[e^k_{\mu}e^i_{\nu} - e^i_{\mu}e^k_{\nu}\right]
\end{equation}
\begin{equation}\label{1.8c}
\de{{\cal L}_G}/de{T^\mu_{ji}} = 4ek\eta_{\mu\sigma}T^{\;\;\;\sigma}_{st}g^{js}g^{it}
\end{equation}
\end{subequations}
In view of this and supposing ${\cal L}_M\/$ independent of torsion and curvature, eqs. \eqref{1.7} become
\begin{subequations}\label{1.9}
\begin{equation}\label{1.9a}
\begin{split}
ee^i_{\mu}\/R - 2e\/R_{\mu\sigma}^{\;\;\;\;\lambda\sigma}e^{i}_\lambda + ek\left(T_{pq}^{\;\;\;\sigma}T^{pq}_{\;\;\;\sigma}e^i_\mu + 4T^{i\;\;\;\sigma}_{\;\;p}T^p_{\;\;k\sigma}e^k_\mu\right) +\\
-\de /de{x^k}\left(4keT^{ki}_{\;\;\;\mu}\right) + 4ke\/\omega_{k\;\;\;\mu}^{\;\;\sigma}T^{ki}_{\;\;\;\sigma}= \de{{\cal L}_M}/de{e^\mu_i}
\end{split}
\end{equation}
and 
\begin{equation}\label{1.9b}
\begin{split}
-\de/de{x^k}\left[2e\/\left(e^k_{\mu}e^i_{\nu} - e^i_{\mu}e^k_{\nu}\right)\right] + \omega_{k\;\;\;\mu}^{\;\;\lambda}\left[2e\/\left(e^k_{\lambda}e^i_{\nu} - e^i_{\lambda}e^k_{\nu}\right)\right]\\
+ \omega_{k\;\;\;\nu}^{\;\;\lambda}\left[2e\/\left(e^k_{\mu}e^i_{\lambda} - e^i_{\mu}e^k_{\lambda}\right)\right] - 4ekT_{\nu\;\;\mu}^{\;\;i} + 4ekT_{\mu\;\;\nu}^{\;\;i}= \de{{\cal L}_M}/de{\omega_i^{\;\;\mu\nu}}
\end{split}
\end{equation}
\end{subequations}
After some calculations, eqs. \eqref{1.9b} may be rewritten in the form 
\begin{equation}\label{3.1.11}
e\/\left( T^{\;\;\;\alpha}_{ts} - T^{\;\;\;\sigma}_{t\sigma}e^\alpha_s + T^{\;\;\;\sigma}_{s\sigma}e^\alpha_t \right)-2ekT_{s\;\;\;t}^{\;\;\alpha} + 2ekT_{t\;\;\;s}^{\;\;\alpha} = \frac{1}{2}\de{{\cal L}_M}/de{\omega_i^{\;\;\mu\nu}}e^\mu_t\/e^\nu_s\/e_i^\alpha
\end{equation}
From \eqref{3.1.11}, by saturating with $e^i_\alpha$, we have
\begin{equation}\label{3.1.12}
T_{ts}^{\;\;\;i} - T_t\delta_s^i + T_s\delta^i_t -2kT_{s\;\;\;t}^{\;\;i} +2kT_{t\;\;\;s}^{\;\;i} =S_{ts}^{\;\;\;i}
\end{equation}
where $S_{ts}^{\;\;\;i} := \frac{1}{2e}\de{{\cal L}_M}/de{\omega_i^{\;\;\mu\nu}}e^\mu_t\/e^\nu_s$ and $T_t :=T^{\;\;\;s}_{ts}$. By contracting the indices $i$ and $s$, we obtain
\begin{equation}\label{3.1.13}
2(k-1)T_t = S_t
\end{equation}
From eq. \eqref{3.1.13} we see that the case $k=1$ is compatible only with $S_t =0$. Supposing then $k\not =1$, we can rewrite eqs. \eqref{3.1.13} in the form
\begin{equation}\label{3.1.14}
T_{ts}^{\;\;\;i} - 2kT_{s\;\;\;t}^{\;\;i} +2kT_{t\;\;\;s}^{\;\;i} = + \frac{1}{2(k-1)}S_t\delta_s^i - \frac{1}{2(k-1)}S_s\delta^i_t + S_{ts}^{\;\;\;i}
\end{equation}
By adding and subtracting the term $2kT_{sti}$ in the left-hand side of eq. \eqref{3.1.14}, we get
\begin{equation}\label{3.1.15}
(1+2k)\/T_{tsi} + 2k\/T_{ist} + 2k\/T_{sti} + 2k\/T_{tis} = + \frac{1}{2(k-1)}S_t\/g_{is} - \frac{1}{2(k-1)}S_s\/g_{it} + S_{tsi}
\end{equation}
On the other hand, by contracting eq. \eqref{3.1.14} with $\epsilon^{htsi}\epsilon_{huvz}$ we get
\begin{equation}\label{3.1.16}
2\/T_{ist} + 2\/T_{sti} + 2\/T_{tis} = \frac{2}{(1-4k)}\left(S_{ist} + S_{sti} + \/S_{tis}\right)
\end{equation}
By inserting the content of \eqref{3.1.16} into eq. \eqref{3.1.15} we end up with the final relation
\begin{equation}\label{3.1.17}
\begin{split}
T_{tsi} = - \frac{2k}{(1-4k)(1+2k)}\left(S_{ist} + S_{sti} + \/S_{tis}\right) + \frac{1}{2(k-1)(1+2k)}S_t\/g_{is} +\\
- \frac{1}{2(k-1)(1+2k)}S_s\/g_{it} + \frac{1}{(1+2k)}S_{tsi}
\end{split}
\end{equation}
where $k \not = \{\frac{1}{4},-\frac{1}{2}\}\/$. Eq. \eqref{3.1.17} expresses the torsion tensor in terms of the spin density. It is worth noticing that in the case $S_{tsi}=0$ we have zero torsion and then the theory reduces to Einstein's General Relativity. We stress that in the previous discussion we have supposed $k \not =\{1,\frac{1}{4},-\frac{1}{2}\}\/$; such values of parameter $k$ are pathological for the present theory and they should be avoided. Indeed, it is easily seen that for $k =\{1,\frac{1}{4},-\frac{1}{2}\}\/$, eqs. \eqref{3.1.12} are unable to uniquely determine the whole torsion tensor. In detail, for $k=1$ eqs. \eqref{3.1.12} become 
\begin{equation}\label{3.1.17.1}
T_{ts}^{\;\;\;i} - T_t\delta_s^i + T_s\delta^i_t -2T_{s\;\;\;t}^{\;\;i} +2T_{t\;\;\;s}^{\;\;i} =S_{ts}^{\;\;\;i}
\end{equation}
and it is a straightforward matter to verify that solutions of \eqref{3.1.17.1} are determined at least up to vector components of the form $V_i\delta^h_j - V_j\delta^h_i\/$, for any $V_i$. Analogously, for $k=\frac{1}{4}$ the torsional equations \eqref{3.1.12} are
\begin{equation}\label{3.1.17.2}
T_{ts}^{\;\;\;i} - T_t\delta_s^i + T_s\delta^i_t - \frac{1}{2}T_{s\;\;\;t}^{\;\;i} + \frac{1}{2}T_{t\;\;\;s}^{\;\;i} =S_{ts}^{\;\;\;i}
\end{equation}
This time, solutions of \eqref{3.1.17.2} are determined at least up to totally antisymmetric components of the form $V_{ijh}=V_{[ijh]}\/$. Finally, for $k=-\frac{1}{2}\/$ eqs. \eqref{3.1.12} assume the form
\begin{equation}\label{3.1.17.3}
T_{ts}^{\;\;\;i} - T_t\delta_s^i + T_s\delta^i_t + T_{s\;\;\;t}^{\;\;i} - T_{t\;\;\;s}^{\;\;i} =S_{ts}^{\;\;\;i}
\end{equation}
and admit solutions defined at least up to traceless non totally antisymmetric components of the kind $V_{ijh} - \frac{1}{3}\left(V_i\/g_{jh} -V_j\/g_{ih}\right) - V_{[ijh]}\/$, where $V_{ijh}= - V_{jih}\/$ is a generic $3$-tensor antisymmetric in the first two indices and $V_i =V_{ij}^{\;\;\;j}$ is its trace vector.

Moreover, by saturating eqs. \eqref{1.9a} with $e^\mu_j$ we get
\begin{equation}\label{3.1.18}
R_{ij} - \frac{1}{2}R\/g_{ij} - \frac{k}{2}\/\left( T_{pqh}T^{pqh}\/g_{ij} + 4T_{ipq}T^{p\;\;\;q}_{\;\;j} \right) + 2k{\tilde\nabla}_h\/T^{h}_{\;\;ij} + 2kK_{hj}^{\;\;\;p}T^{h}_{\;\;ip} = \Sigma_{ij}
\end{equation}
where, using the relationships $R^h_{\;\;kij}=R_{ij}^{\;\;\;\;\mu\sigma}\eta_{\sigma\nu}e^h_\mu e^\nu_k\/$ and $T_{ij}^{\;\;\;h}=T^{\;\;\;\mu}_{ij}e_\mu^h\/$ among the curvature and torsion tensors related respectively to the spin connection $\omega_i^{\;\;\mu\nu}\/$ and the associated linear connection $\Gamma_{ij}^{\;\;\;h}=e^h_\mu\left(\de{e^\mu_j}/de{x^i} + \omega_{i\;\;\;\nu}^{\;\;\mu}e^\nu_j\right)\/$, $R_{ij}:=R^h_{\;\;ihj}$ and $R=R^i_{\;\;i}$ are the Ricci tensor and scalar curvature associated with the linear connection $\Gamma$, ${\tilde\nabla}_h$ denotes the Levi--Civita covariant derivative associated with the metric tensor $g_{ij}$, $\Sigma^i_{\;j}:=- \frac{1}{2e}\de{{\cal L}_M}/de{e^\mu_i}\/e^\mu_j$ indicates the matter energy-impulse tensor and 
\begin{equation}\label{3.1.19}
K_{ij}^{\;\;\;h}
=\frac{1}{2}\left(-T_{ij}^{\;\;\;h}+T_{j\;\;\;i}^{\;\;h}-T^{h}_{\;\;ij}\right)
\end{equation}
represents the usual contorsion tensor.Recalling the decomposition of the metric compatible connection $\Gamma$ in terms of the Levi--Civita connection $\tilde\Gamma$ and contorsion tensor $K$
\begin{equation}\label{3.1.20}
\Gamma_{ij}^{\;\;\;h}=\tilde{\Gamma}_{ij}^{\;\;\;h}-K_{ij}^{\;\;\;h}
\end{equation}
we can decompose the Ricci tensor as
\begin{equation}\label{3.1.21}
R_{ij}=\tilde{R}_{ij} + \tilde{\nabla}_jK_{hi}^{\;\;\;h} - \tilde{\nabla}_h\/K_{ji}^{\;\;\;h} + K_{ji}^{\;\;\;p}K_{hp}^{\;\;\;h} - K_{hi}^{\;\;\;p}K_{jp}^{\;\;\;h}
\end{equation}
and then, inserting the content of eqs. \eqref{3.1.19} and \eqref{3.1.20} into eqs. \eqref{3.1.18}, we can separate the Levi--Civita contributions from the torsional ones. 

To conclude this section, we recall the conservation laws holding for energy and spin
\begin{equation}\label{5.15}
\nabla_i\/\Sigma^{ij} + T_i\/\Sigma^{ij} - \Sigma_{is}T^{jis} - \frac{1}{2}R^{tsij}S_{tsi} = 0
\end{equation}
\begin{equation}\label{5.11}
\nabla_h\/S^{tsh} + T_h\/S^{tsh} + \Sigma^{ts} - \Sigma^{st} = 0
\end{equation}
The latter result from the invariance of the Lagrangian under diffeomorphisms and Lorentz transformations (see, for example, \cite{Poplawski}). Alternatively, eqs. \eqref{5.15} and \eqref{5.11} can be derived directly form the field equations making use of the Bianchi identities. 
\section{The Coupling to Matter Fields}
\label{4}
\subsection{The Coupling to the Dirac Spinor}
In this section we study the above theory coupled to a Dirac field. The matter Lagrangian of the Dirac field will be taken without any modification with respect to the standard one, the only adjustment will be that now, spinorial covariant derivatives will be the most general containing torsion, as it was first discussed by Fock and Ivanenko, and as also reported  for instance in \cite{Hehl:1994ue}: the Lagrangian is thus given by
\begin{equation}
\label{4.1}
{\cal L}_M=
\left[\frac{i}{2}\left(\bar\psi\gamma^iD_{i}\psi-D_{i}\bar\psi\gamma^{i}\psi\right)-m\bar\psi\psi\right]
\end{equation}
where $D_i\psi = \de\psi/de{x^i} + \omega_i^{\;\;\mu\nu}S_{\mu\nu}\psi\/$ and $D_i\bar\psi = \de{\bar\psi}/de{x^i} - \bar\psi\omega_i^{\;\;\mu\nu}S_{\mu\nu}\/$ are the covariant derivatives of the Dirac fields, $S_{\mu\nu}=\frac{1}{8}\left[\gamma_\mu,\gamma_\nu\right]\/$, $\gamma^i =\gamma^{\mu}e^i_\mu\/$ with $\gamma^\mu\/$ denoting Dirac matrices and where $m$ is the mass of the Dirac field. From eq. \eqref{4.1}, it is easily seen that the spin density tensor, defined as in section 3, is given by
\begin{equation}\label{4.2b}
S_{ij}^{\;\;\;h}=\frac{i}{2}\bar\psi\left\{\gamma^{h},S_{ij}\right\}\psi
\equiv-\frac{1}{4}\eta^{\mu\sigma}\epsilon_{\sigma\nu\lambda\tau}
\left(\bar{\psi}\gamma_{5}\gamma^{\tau}\psi\right)e^{h}_{\mu}e^{\nu}_{i}e^{\lambda}_{j}
\end{equation}
In this case, being the spin density tensor \eqref{4.2b} totally antisymmetric, the torsion tensor \eqref{3.1.17} assumes the explicit expression
\begin{equation}\label{4.3}
T_{tsi}=\kappa\/S_{tsi}
\end{equation} 
where $\kappa=\frac{1}{1-4k}$ is a constant depending on $k$ ($k\not = \frac{1}{4}$). Moreover, from eq. \eqref{3.1.19} we have
\begin{equation}\label{4.4}
K_{ij}^{\;\;\;p} = -\frac{\kappa}{2}S_{ij}^{\;\;\;p}
\end{equation}
By varying \eqref{4.1} with respect to $\psi$, we obtain the Dirac equations
\begin{equation}\label{4.4bis}
i\gamma^h\/D_h\psi - m\psi =0
\end{equation}
Due to eqs. \eqref{4.4bis}, it is easily seen that the energy--impulse tensor, derived as in section 3, is expressed as 
\begin{equation}
\label{4.2a}
\Sigma_{ij}=\frac{i}{4}\left(\bar\psi\gamma_{i}D_{j}\psi-D_{j}\bar\psi\gamma_{i}\psi\right)
\end{equation}
The Dirac equations \eqref{4.4bis} imply the validity of the identities (see, for example, \cite{FV})
\begin{subequations}\label{4.4tris}
\begin{equation}\label{4.4trisa}
\nabla_i\Sigma^{ij}=T^{jik}\Sigma_{ik} + \frac{1}{2}S_{pqi}R^{pqij}
\end{equation}
\begin{equation}\label{4.4trisb}
\nabla_h\/S^{ijh} = \tilde{\nabla}_h\/S^{ijh} = \Sigma^{ji} - \Sigma^{ij}
\end{equation}
\end{subequations}
where the total antisymmetry of the spin and torsion has been systematically used. In addition to this, inserting \eqref{4.4} into \eqref{3.1.21} we obtain the decomposition of the Ricci tensor and scalar curvature 
\begin{subequations}\label{4.5}
\begin{equation}\label{4.5a}
R_{ij}=\tilde{R}_{ij} + \frac{\kappa}{2}\/\tilde{\nabla}_p\/S_{ji}^{\;\;\;p} - \frac{\kappa^2}{4}S_{pi}^{\;\;\;q}\/S_{jq}^{\;\;\;p}
\end{equation}
\begin{equation}\label{4.5b}
R=\tilde{R} - \frac{\kappa^2}{4}\/S_{qpr}\/S^{qpr}
\end{equation}
\end{subequations}
as well as the identities
\begin{subequations}\label{4.6}
\begin{equation}\label{4.6a}
\frac{1}{2}\/\left( T_{pqh}T^{pqh}\/g_{ij} + 4T_{ipq}T^{p\;\;\;q}_{\;\;j} \right) =-\left( \frac{\kappa^2}{2}S_{pqh}S^{pqh}\/g_{ij} - 2\kappa^2\/S_{piq}S_j^{\;\;qp} \right)
\end{equation}
\begin{equation}\label{4.6b}
2{\tilde\nabla}_p\/T^{p}_{\;\;ij} = 2\kappa{\tilde\nabla}_p\/S^{p}_{\;\;ij}
\end{equation}
\begin{equation}\label{4.6c}
2K_{hj}^{\;\;\;p}T^{h}_{\;\;ip} = - \kappa^2S_{pj}^{\;\;\;q}S^{p}_{\;\;iq}
\end{equation}
\end{subequations}
In view of eqs. \eqref{4.5} and \eqref{4.6}, we can rewrite eqs. \eqref{3.1.18} in the form
\begin{equation}\label{4.7}
\begin{split}
\tilde{R}_{ij}-\frac{1}{2}\tilde{R}g_{ij}+\frac{\kappa}{2}\/\tilde{\nabla}_p\/S_{ji}^{\;\;\;p} - \frac{\kappa^2}{4}S_{pi}^{\;\;\;q}\/S_{jq}^{\;\;\;p}+\frac{\kappa^2}{8}\/S_{qpr}\/S^{qpr}g_{ij} +\\
- k\kappa^2\left( \frac{1}{2}S_{pqh}S^{pqh}\/g_{ij}-2S_{piq}S_j^{\;\;qp} \right) + 2\kappa k{\tilde\nabla}_p\/S^{p}_{\;\;ij} - \kappa^2k\/S_{pj}^{\;\;\;q}S^{p}_{\;\;iq} = \Sigma_{ij}
\end{split}
\end{equation}
We can decompose eqs. \eqref{4.7} in their symmetric and antisymmetric parts, that is
\begin{subequations}\label{4.8}
\begin{equation}\label{4.8a}
\begin{split}
\tilde{R}_{ij} - \frac{1}{2}\tilde{R}\/g_{ij}-\frac{\kappa^2}{4}S_{pi}^{\;\;\;q}\/S_{jq}^{\;\;\;p} + \frac{\kappa^2}{8}\/S_{qpr}\/S^{qpr}g_{ij}-k\left(\frac{\kappa^2}{2}S_{pqh}S^{pqh}\/g_{ij}- 2\kappa^2\/S_{piq}S_j^{\;\;qp} \right) +\\
- \kappa^2k\/S_{pj}^{\;\;\;q}S^{p}_{\;\;iq} = \Sigma_{(ij)}
\end{split}
\end{equation}
\begin{equation}\label{4.8b}
\frac{\kappa}{2}\/\tilde{\nabla}_p\/S_{ji}^{\;\;\;p} + 2\kappa k{\tilde\nabla}_p\/S^{p}_{\;\;ij} = \Sigma_{[ij]}
\end{equation}
\end{subequations}
After some algebraic calculations, eqs. \eqref{4.8} simplify as
\begin{subequations}\label{4.8bis}
\begin{equation}\label{4.8bisa}
\tilde{R}_{ij} - \frac{1}{2}\tilde{R}\/g_{ij} - \frac{\kappa}{4}\/S_{pi}^{\;\;\;q}\/S_{jq}^{\;\;\;p} + \frac{\kappa}{8}\/S_{qpr}\/S^{qpr}g_{ij} = \Sigma_{(ij)}
\end{equation}
\begin{equation}\label{4.8bisb}
\frac{1}{2}\/\tilde{\nabla}_p\/S_{ji}^{\;\;\;p} = \Sigma_{[ij]}
\end{equation}
\end{subequations}
Eqs. \eqref{4.8bisb} are clearly identical to eqs. \eqref{4.4trisb}, automatically ensured by the Dirac equations themselves. This means that the significant part of the Einstein--like equations \eqref{4.7} reduces to the symmetric one \eqref{4.8bisa}. The latter can be worked out by giving an explicit representation of the covariant derivative of spinor fields. To this end, we denote by $\tilde D$ the spinorial covariant derivative induced by the Levi--Civita connection. Then
\begin{subequations}\label{4.8tris}
\begin{equation}\label{4.8trisa}
D_i\/\psi = \tilde{D}_i\/\psi - \frac{1}{4}K_{ijh}\gamma^h\gamma^j\psi
\end{equation}
\begin{equation}\label{4.8trisb}
D_i\/\bar{\psi} = \tilde{D}_i\/\bar{\psi} + \frac{1}{4}\bar{\psi}K_{ijh}\gamma^h\gamma^j
\end{equation}
\end{subequations}
Making use of eqs. \eqref{4.2b}, \eqref{4.4} and \eqref{4.8tris}, the following identities are easily verified
\begin{subequations}\label{4.9.0}
\begin{equation}\label{4.9}
\begin{split}
\bar\psi\gamma_iD_j\psi - \left(D_j\bar\psi\right)\gamma_i\psi = \bar\psi\gamma_i\tilde{D}_j\psi - (\tilde{D}_j\bar\psi)\gamma_i\psi -\frac{i\kappa}{8}(\bar{\psi}\gamma_5\gamma^\tau\psi)(\bar{\psi}\gamma_5\gamma_\tau\psi)g_{ij} \\
+ \frac{i\kappa}{8}(\bar{\psi}\gamma_5\gamma_i\psi)(\bar{\psi}\gamma_5\gamma_j\psi) 
\end{split}
\end{equation}
\begin{equation}\label{4.10}
S_{hi}^{\;\;\;p}S_{jp}^{\;\;\;h} = - \frac{1}{8}\left(\bar\psi\gamma_5\gamma^\tau\psi\right)\left(\bar\psi\gamma_5\gamma_\tau\psi\right)g_{ij} + \frac{1}{8}\left(\bar\psi\gamma_5\gamma_i\psi\right)\left(\bar\psi\gamma_5\gamma_j\psi\right)
\end{equation} 
\begin{equation}\label{4.11}
S_{hqp}S^{hqp} = - \frac{3}{8}\left(\bar\psi\gamma_5\gamma^\tau\psi\right)\/\left(\bar\psi\gamma_5\gamma_\tau\psi\right)
\end{equation}
\end{subequations}
Inserting the content of eqs. \eqref{4.9.0} into eqs. \eqref{4.8bisa}, we end up with final symmetric Einstein--like equations of the form
\begin{equation}\label{4.12}
\tilde{R}_{ij} - \frac{1}{2}\tilde{R}\/g_{ij} = \frac{i}{4}\left[\bar\psi\gamma_{(i}\tilde{D}_{j)}\psi - (\tilde{D}_{(j}\bar\psi)\gamma_{i)}\psi\right] + \frac{3\kappa}{64}\left(\bar\psi\gamma_5\gamma^\tau\psi\right)\left(\bar\psi\gamma_5\gamma_\tau\psi\right)g_{ij}
\end{equation}
It is worth noticing that the conservation laws for the energy \eqref{4.4trisa} reduce to the simpler
\begin{equation}\label{4.14}
\tilde{\nabla}_i\tilde{\Sigma}^{(ij)} -\frac{\kappa}{8}\tilde{\nabla}^j\left(S_{qih}S^{qih}\right) =0
\end{equation}
where $\tilde{\Sigma}_{ij} := \frac{i}{4}\left[\bar\psi\gamma_{i}\tilde{D}_{j}\psi - (\tilde{D}_{j}\bar\psi)\gamma_{i}\psi\right]$. Comparing with eqs. \eqref{4.11} and \eqref{4.12}, eqs. \eqref{4.14} ensure that the Levi--Civita quadridivergence of the effective energy--impulse tensor on the right hand side of \eqref{4.12} vanishes.

At the same time, the Dirac equations can be expressed as
\begin{equation}\label{4.13}
i\gamma^i\tilde{D}_i\psi-\lambda\bar{\psi}\gamma^{q}\psi\gamma_q\psi-m\psi=0
\end{equation}
where $\lambda:=\frac{3}{16(1-4k)}\equiv\frac{3\kappa}{16}$ from now on. Eqs. \eqref{4.13} can straightforwardly be cast after a Fierz rearrangement into an equivalent but simpler form given by
\begin{equation}
\label{Diracreduced}
i\gamma^i\tilde{D}_i\psi
+\lambda\left(\bar{\psi}\gamma_{5}\psi\gamma_5-\bar{\psi}\psi\right)\psi-m\psi=0
\end{equation}
formally identical to what we would have had if it were in the torsionless case but with additional potentials of self-interactions. 

Notice that since in our approach generalizations have place only for the geometrical sector, and not for the material sector, then we obtain a generalized form of the gravitational (metric and torsion) field equations, but not of the matter (Dirac) field equations, which remain formally the same we would have had as usual \cite{Hehl:1971qi,Mielke:2004gg}; however, since the torsion-spin density field equations are now different, so soon as torsion is replaced through these field equations with the spin density of the spinor, the eventual non-linear interaction in the spinorial field equation (the self-interaction of fermions) assumes a form that is in general different from that of \cite{Hehl:1971qi}: although due to the peculiar structure of the Dirac spinor these non-linear fermionic interactions in the matter field equations do happen to be reduced to the usual \emph{form}, nevertheless this does not happen for their \emph{strength} as the coupling constant in now different. When in gravity torsion is not neglected, non-linearities as those given in \cite{Hehl:1971qi} appear within the Dirac matter field equations, but when the present modification of the torsional action is accounted, those non-linearities get a coupling constant that is not necessarily the Newton constant: so it may seem that our modification has the only effect of rescaling the torsional coupling constant and nothing more. However, in a situation in which one of the most important effects of torsion is its induced spin-contact coupling in the matter field equations and in a moment in which torsion tends to be neglected because such interactions are allegedly small, our modification, far from being a mere rescaling of the torsion coupling constant, is what renders those interactions relevant at scales in which torsion has always been assumed to be negligible, solving the last impeachment torsion would have to face. The possibility to control the torsional coupling constant gives one the opportunity to change its sign as well, making the torsionally-induced interactions not only important at larger scales, but also attractive.

In fact, in the present theory as in the Einstein-Cartan theory, the Dirac field equations are of the Nambu-Jona--Lasinio type \cite{Fabbri:2010rw}; however, in this model the Dirac field has self-interactions whose coupling constant can be taken with opposite sign, giving rise to Nambu-Jona--Lasinio attractive potentials \cite{Fabbri:2011kq}: in the low-energy limit, this gives the possibility to describe superconductivity, therefore providing a geometric interpretation.

Since our modification affects only the value of the coupling constant, but not the structure of the non-linearities, then our model has all advantages but also all disadvantages of the usual model of Dirac theory with torsion: for instance, the issue of chiral anomalies \cite{Mielke:2006gi} and the fact that the non-linearities are perturbatively non-renormalizable \cite{Eichhorn:2011pc}; for these, and maybe others, among all the problems the theory may have we furnish no solution more than those discussed in \cite{Mielke:2006gi,Eichhorn:2011pc}, or in other works, and therefore our modification is neither better, nor worse, than the usual theory discussed in the above-mentioned references. By this we do not mean that these problems are not important, and on the contrary they ought be analyzed; what we only mean is that these issues are untouched by the modification propounded by our work, and so we are not obliged to discuss them more than any other paper on torsion for Dirac fields. Our paper is focused on the scaling of their coupling constant, and this is the key point we want to deepen next.

It is important to notice that no matter what this value is, it would at first be thinkable to normalize it to the usual value of the Newton constant through the normalization of the Dirac field, but a deeper analysis shows that this process would also have the effect of changing the scale of the energy density within the gravitational field equations and it is therefore unacceptable: so far as our knowledge is concerned, the theory we have here constructed is the only one in which the Einstein-Cartan-Dirac theory can be generalized in order for the system of field equations, once decomposed in terms of torsionless quantities plus torsional contributions written as Dirac field self-interactions, to provide a coupling constant for self-interactions of matter leaving gravity unmodified.

\paragraph{1. Neutrino oscillation in absence of mass states.} The first example we will consider in the case of a coupled system of fermions is given by the simplest couple of spinors, that is the pair of semispinors both massless; these semispinors without mass will possess spin-torsion coupling providing some sort of mixing between them despite the fact that in the accepted model neutrinos only have massive oscillations: the aim is therefore to compare this case with the accepted model to appreciate their discrepancies.

Considering the field equation for two semispinors without mass $\nu_{1}$ and $\nu_{2}$ we have that their field equations are given by the usual field equations for semispinors taken to be left-handed and so in the purely massless configuration
\begin{eqnarray}
&i\gamma^{\mu}D_{\mu}\nu_{1}=0\\
&i\gamma^{\mu}D_{\mu}\nu_{2}=0
\end{eqnarray}
in terms of the most general covariant derivatives $D_{\mu}$ that can be decomposed into the simplest covariant derivatives $\tilde{D}_{\mu}$ as in the torsionless case plus torsional contributions as
\begin{eqnarray}
&i\gamma^{\mu}\tilde{D}_{\mu}\nu_{1}
+\lambda\bar{\nu}_{2}\gamma_{\mu}\nu_{2}\gamma^{\mu}\nu_{1}=0\\
&i\gamma^{\mu}\tilde{D}_{\mu}\nu_{2}
+\lambda\bar{\nu}_{1}\gamma_{\mu}\nu_{1}\gamma^{\mu}\nu_{2}=0
\end{eqnarray}
with spinorial interactions of each semispinor with the other; these have now to be written in an alternative form by employing Fierz identities $\bar{\psi}\gamma_{\mu}\psi\gamma^{\mu}\chi
=-\bar{\psi}\gamma_{\mu}\chi\gamma^{\mu}\psi$ valid for any couple of left-handed semispinors $\psi$ and $\chi$ identically: after this Fierz rearrangement we get the previous field equations transcribed as 
\begin{eqnarray}
&i\gamma^{\mu}\tilde{D}_{\mu}\nu_{1}
-\frac{\lambda}{3}\left(\bar{\nu}_{1}\gamma_{\mu}\nu_{1}
-\bar{\nu}_{2}\gamma_{\mu}\nu_{2}\right)\gamma^{\mu}\nu_{1}
-\frac{2\lambda}{3}\left(\bar{\nu}_{2}\gamma^{\mu}\nu_{1}\right)\gamma_{\mu}\nu_{2}=0\\
&i\gamma^{\mu}\tilde{D}_{\mu}\nu_{2}
-\frac{2\lambda}{3}\left(\bar{\nu}_{1}\gamma^{\mu}\nu_{2}\right)\gamma_{\mu}\nu_{1}
+\frac{\lambda}{3}\left(\bar{\nu}_{1}\gamma_{\mu}\nu_{1}
-\bar{\nu}_{2}\gamma_{\mu}\nu_{2}\right)\gamma^{\mu}\nu_{2}=0
\label{fieldequations}
\end{eqnarray}
in which the interactions of each semispinor with the other has been written in a form that is particularly interesting; in fact, such a coupled system of field equations is formally
\begin{eqnarray}
&\!\!\!\!i\gamma^{\mu}\!
\left[\tilde{D}_{\mu}\!\!
\left(\!\!\!\!\begin{tabular}{c}$\nu_{1}$\\$\nu_{2}$\end{tabular}\!\!\!\!\right)
\!+\!\frac{i\lambda}{3}\!\left(\!\!\!\!\begin{tabular}{cc}
$(\bar{\nu}_{1}\gamma_{\mu}\nu_{1}-\bar{\nu}_{2}\gamma_{\mu}\nu_{2})$ & $2(\bar{\nu}_{1}\gamma_{\mu}\nu_{2})^{\ast}$\\ $2(\bar{\nu}_{1}\gamma_{\mu}\nu_{2})$ & $-(\bar{\nu}_{1}\gamma_{\mu}\nu_{1}-\bar{\nu}_{2}\gamma_{\mu}\nu_{2})$ \end{tabular}\!\!\!\!\right)\!\!
\left(\!\!\!\!\begin{tabular}{c}$\nu_{1}$\\ $\nu_{2}$\end{tabular}\!\!\!\!\right)\!\right]\!=\!0
\end{eqnarray}
in which we achieve a considerable compactification. Next we simply have to call the doublet of semispinors
\begin{eqnarray}
&\left(\!\!\begin{tabular}{c}$\nu_{1}$\\$\nu_{2}$\end{tabular}\!\!\right)=\nu
\end{eqnarray}
with which to define the triplet of vectors
\begin{eqnarray}
&\frac{1}{3}\bar{\nu}\gamma_{\mu}\vec{\sigma}\nu=\vec{A}_{\mu}
\end{eqnarray}
so that we have
\begin{eqnarray}
&\!\!\!\!i\gamma^{\mu}\!
\left[\tilde{D}_{\mu}\nu
+i\lambda\left(\!\!\!\!\begin{tabular}{cc}
$A_{\mu}^{3}$ & $A_{\mu}^{1}-iA_{\mu}^{2}$\\ $A_{\mu}^{1}+iA_{\mu}^{2}$ & $-A_{\mu}^{3}$ \end{tabular}\!\!\!\!\right)\nu\right]\!=\!0
\end{eqnarray}
or equivalently
\begin{eqnarray}
&i\gamma^{\mu}\left[\tilde{D}_{\mu}\nu+i\lambda\vec{A}_{\mu}\cdot\vec{\sigma}\nu\right]=0
\end{eqnarray}
as it can be checked; then upon introduction of 
\begin{eqnarray}
&\mathbb{D}_{\mu}\nu=\tilde{D}_{\mu}\nu+2i\lambda\vec{A}_{\mu}\cdot\frac{\vec{\sigma}}{2}\nu
\label{derivative}
\end{eqnarray}
we finally have
\begin{eqnarray}
&i\gamma^{\mu}\mathbb{D}_{\mu}\nu=0
\label{neutrinos}
\end{eqnarray}
in which the free covariant derivative $\tilde{D}_{\mu}$ is now part of a gauge covariant derivative $\mathbb{D}_{\mu}$, and thus this is the field equation in the torsionless case but in presence of an additional gauge interaction. To see that this is actually a gauge invariant interaction, notice that as the doublet of semispinors $\nu$ transforms according to the $SU(2)$ group 
\begin{eqnarray}
&\nu'=e^{2i\lambda\vec{\theta}\cdot\frac{\vec{\sigma}}{2}}\nu
\end{eqnarray}
then the triplet of vectors $\vec{A}_{\mu}$ and the triplet of Pauli matrices that generate the infinitesimal transformation of the $SU(2)$ group are combined into $\vec{A}_{\mu}\cdot\vec{\sigma}$ transforming as 
\begin{eqnarray}
&\left[\vec{A}_{\mu}\cdot\frac{\vec{\sigma}}{2}\right]'
=e^{2i\lambda\vec{\theta}\cdot\frac{\vec{\sigma}}{2}}
\left[\left(\vec{A}_{\mu}-\partial_{\mu}\vec{\theta}\right)\cdot\frac{\vec{\sigma}}{2}\right]
e^{-2i\lambda\vec{\theta}\cdot\frac{\vec{\sigma}}{2}}
\end{eqnarray}
as the gauge connection of the $SU(2)$ group and field equations \eqref{neutrinos} are invariant for $SU(2)$ local transformations \cite{F/1}. Here $\lambda$ is to be tuned to the neutrinos' oscillation length.

The field equations \eqref{neutrinos} are still the field equations for semispinors in the massless configuration but now due to the masslessness of the neutrinos their covariant derivative contains interactions gauging the $SU(2)$ group: so that the two types of neutrinos will be converted into one another according to a sort of massless oscillation \cite{F/1}. This is a remarkable discrepancy between the present theory, in which it is precisely because neutrinos are massless that they may oscillate, and the commonly accepted theory, in which only massive neutrinos can possibly oscillate.

\paragraph{2. Leptons with weak interactions and Higgs field.} Our next example will be the immediately more complex one, that is the coupled system of fermions, one of which is a spinor with mass while the other is a semispinor thus massless; these spinors with massive states will possess spin-torsion coupling resulting in interactions between the two fundamental fields while from the point of view of the standard model the couple of electron and neutrino only has the weak interactions: the aim is therefore to compare this case with that of the weak forces and see whether or not similarities may arise. 

In this case the field equations for the electron and neutrino fields $e$ and $\nu$ are given by the previous field equations, but because now the system is constituted by a couple of spinors then the spin will be the sum of the two spins and the spinorial field equations are
\begin{eqnarray}
&i\gamma^{\mu}\tilde D_{\mu} e
-\lambda\left(\overline{e}\gamma_{\mu}e\gamma^{\mu}e
+\overline{\nu}\gamma_{\mu}\nu\gamma^{\mu}\gamma_{5}e\right)-me=0\\
&i\gamma^{\mu}\tilde D_{\mu}\nu
-\lambda\overline{e}\gamma_{\mu}\gamma_{5}e\gamma^{\mu}\nu=0
\end{eqnarray}
in which the fact that the electron is massive while the neutrino is massless is the reason that prevents these two fields to mix into a doublet; after by employing a Fierz rearrangement as we have done before, they can be cast into the form
\begin{eqnarray}
\label{equations}
\nonumber
&i\gamma^{\mu}\tilde D_{\mu}e+2\lambda(\cos{\theta})^{2}\overline{e}\gamma_{5}e\gamma_{5}e+\\
&+q\tan{\theta}Z_{\mu}\gamma^{\mu}e-\frac{g}{2\cos{\theta}}Z_{\mu}\gamma^{\mu}e_{L}
+\frac{g}{\sqrt{2}}W^{*}_{\mu}\gamma^{\mu}\nu-He-me=0\label{electron}\\
&i\gamma^{\mu}\tilde D_{\mu}\nu+\frac{g}{2\cos{\theta}}Z_{\mu}\gamma^{\mu}\nu
+\frac{g}{\sqrt{2}}W_{\mu}\gamma^{\mu}e_{L}=0
\label{neutrino}
\end{eqnarray}
once we define
\begin{eqnarray}
&Z^{\mu}
=-\lambda\left[2(\sin{\theta})^{2}\overline{e}\gamma^{\mu}e-\overline{e}_{L}\gamma^{\mu}e_{L}
+\overline{\nu}\gamma^{\mu}\nu\right]\left(\frac{\cot{\theta}}{q}\right)
\label{neutral}\\
&W^{\mu}
=-\lambda\left(\overline{e}_{L}\gamma^{\mu}\nu\right)
\left[\frac{4(\sin{\theta})^{2}-1}{q\sqrt{2}\sin{\theta}}\right]
\label{charged}
\end{eqnarray}
and 
\begin{eqnarray}
&H=\lambda\overline{e}e2(\cos{\theta})^{2}
\label{Higgs}
\end{eqnarray}
showing that field equations (\ref{electron}) and (\ref{neutrino}) are formally identical to the system of field equations for the lepton fields after the symmetry breaking in the standard model, although both weak and Higgs boson fields are here composite \cite{f/1}. Notice that because the field equations of the leptonic fields are known, it is possible to compute the divergence of the leptonic currents (\ref{neutral}-\ref{charged}), for which at the weak interaction scales a reasonable estimate gives the partially conserved axial currents
\begin{eqnarray}
&\frac{gv}{\sqrt{2}\cos{\theta}}
\left[\left(1\!+\!\frac{H}{v}\right)\!\nabla_{\mu}Z^{\mu}
\!+\!2Z^{\mu}\nabla_{\mu}\frac{H}{v}\right]=-\frac{mi\overline{e}\gamma e}{v\sqrt{2}}
\label{conservedneutral}\\
&\frac{gv}{\sqrt{2}}\left[\!\left(1+\!\frac{H}{v}\right)\!\left(\nabla_{\mu}W^{\mu}\!+\!iq\tan{\theta}Z^{\mu}W_{\mu}\right)
\!+\!2W^{\mu}\nabla_{\mu}\frac{H}{v}\right]=\frac{mi\overline{e}\gamma\nu}{v}
\label{conservedcharged}
\end{eqnarray}
which in the limit of the approximation are precisely the partially conserved axial currents one would have had in the standard model so soon as the torsional coupling constant is tuned as $4\lambda=\frac{1}{v^{2}}$ in terms of the Higgs vacuum \cite{Fabbri:2012zd}. Notice that with this fine-tuning for the constant, the weak interactions are reproduced not only in structure but also in strength while the masses of the composite mediators have not only the proper ratio but also the correct value. This would imply that among theories coming from torsion and the standard model, so long as we take into account leptonic weak scattering and the weak mediators masses it is not possible to appreciate any discrepancies, and we have to go to higher energies to see that in our model the weak mediators must display an internal structure that can never be present for the standard model weak bosons \cite{Fabbri:2009ta}.

Similar results have also been obtained both for leptons and hadrons in circumstances in which the underlying symmetry was still unbroken \cite{f/2,f/3}.

\paragraph{3. Dirac field as a Van der Waals gas.} We will now leave the treatment of the coupled system of Dirac fields to focus on the single Dirac field; writing the Dirac field equation in the standard representation allows us to obtain the slow-speed weak-field approximation
\begin{eqnarray}
&i\frac{\partial\phi}{\partial t}
+\frac{1}{2m}\boldsymbol{\sigma}^{k}\boldsymbol{\tilde{D}}_{k}
\boldsymbol{\sigma}^{a}\boldsymbol{\tilde{D}}_{a}\phi
-\lambda\left(\phi^{\dagger}\phi\right)\phi-m\phi=0
\label{matterfieldapproximated}
\end{eqnarray}
representing cold-matter fields with self-interactions with the structure of an energy term.

Thus in the present theory we have that the Dirac field equations have the form of the Nambu-Jona--Lasinio field equation further approximated to the Pauli-Schr\"{o}dinger field equation of the Ginzburg-Landau type \cite{Fabbri:2010rw}. In it the coupling constant $\lambda$ is to be set on the value of the specific condensed state system we would eventually like to study \cite{Fabbri:2011kq}.

These field equations have energy levels that can easily be computed to be
\begin{eqnarray}
&E=m+\frac{p^{2}}{2m}+\lambda u^{2}
\label{energy}
\end{eqnarray}
where $u^{2}$ is the square of the field, representing the density of the field itself, or equivalently in the quantum-mechanical interpretation the probability to find a give particle or again for collective states the number of particles per unit volume $u^{2}=\frac{N}{V}$; on the other hand, the usual thermodynamic interpretation of the kinetic energy is the temperature $T$ of the gas under consideration: considering both these interpretations, the energy levels \eqref{energy} receive the quantum-mechanical thermodynamic interpretation of 
\begin{eqnarray}
&E=T+\lambda \frac{N}{V}
\label{energyvdw}
\end{eqnarray} 
which is the well known expression for the energy level of the Van der Waals gas.

By employing the equation of the energy $\left(\frac{\partial E}{\partial V}\right)_{T}= \left(\frac{\partial P}{\partial T}\right)_{V}T-P$ it is easy to see that the equation of state gives the pressure $P$ as a function of volume and temperature as
\begin{eqnarray}
&\left(P-\lambda\frac{N}{V^{2}}\right)F(V)=T
\label{equationstate}
\end{eqnarray}
as the Van der Waals equation of state with generalized volume factor $F(V)$ and with coupling constant $\lambda$ determining the type of interaction of the gas: for the commonly accepted positive values the Van der Waals pressure is positive resulting into a repulsion, while for negative values the Van der Waals pressure is negative resulting into an attraction. 

The interpretation of a Dirac field or a collective state of Dirac particles as a Van der Waals gas can be read by interpreting the torsionally-induced self-interactions as the Van der Waals interactions among different particles of the gas, as it might have been expected from the fact that both type of interactions drop with a $\frac{1}{r^{6}}$ dependence \cite{sa-si}.
\subsection{The Coupling to the Fluid with Spin}
The importance of the Dirac field is mainly due to the fact that it is the simplest quantum field that can be used in the description of natural phenomena, including even cosmological applications (see for instance \cite{Mielke:2006zp}); other spin-$\frac{1}{2}$ spinor fields like ELKO may also be used in cosmology such as those tackling the problem of Dark Matter \cite{Fabbri:2012yg}. In the present paper however, we wish to employ yet another matter field to describe large-scale structures: the fluid with spin, that is the Weyssenhoff fluid \cite{Obukhov,Hehl-Heyde-Kerlick}. 

The spin fluid is defined by its energy-momentum tensor, of the form
\begin{subequations}\label{6.1}
\begin{equation}\label{6.1a}
\Sigma^{ij}= U^iP^j + p\left( U^iU^j - g^{ij}\right)
\end{equation}
and its spin density tensor given by
\begin{equation}\label{6.1b}
S_{ij}^{\;\;\;h}=S_{ij}U^h
\end{equation}
\end{subequations}
where $U^i $ ($U^iU_i=1 $) and $P^i$ denote respectively the $4$-velocity and the $4$-vector density of energy-momentum, while $S_{ij}=-S_{ji}$ is the spin density of the fluid: the $4$-velocity and the spin satisfy the convective condition
\begin{equation}\label{6.2}
S_{ij}U^j =0.
\end{equation}

Making use of eqs. \eqref{3.1.19}, \eqref{6.1b} and \eqref{6.2}, we obtain the expression of the contorsion
\begin{equation}\label{6.3}
K_{tsi}=\frac{1}{2}\left[-A(k) \left(S_{sti}+S_{tis}+S_{ist}\right) + B(k) \left(-S_{tsi}+S_{sit}-S_{its}\right)\right]
\end{equation}
where
\begin{equation}\label{6.4}
A(k) = \frac{-2k}{(1-4k)(1+2k)} \qquad {\rm and} \qquad B(k) = \frac{1}{(1+2k)}
\end{equation}
and due to the convective condition \eqref{6.2}, it is seen that torsion and contorsion tensors are both traceless. From this, we can write the Einstein-like equations in the form
\begin{equation}\label{6.5}
\begin{split}
\tilde{R}_{ij} - \frac{1}{2}\tilde{R} g_{ij} - \tilde{\nabla}_h K_{ji}^{\;\;\;h} - K_{hi}^{\;\;\;p}K_{jp}^{\;\;\;h} +\frac{1}{2}K_{h}^{\;\;ip}K_{ip}^{\;\;\;h}g_{ij} +\\
- \frac{k}{2} \left( T_{pqh}T^{pqh} g_{ij} + 4T_{ipq}T^{p\;\;\;q}_{\;\;j} \right) + 2k{\tilde\nabla}_h T^{h}_{\;\;ij} + 2kK_{hj}^{\;\;\;p}T^{h}_{\;\;ip} = \Sigma_{ij}
\end{split}
\end{equation}
It is easy to verify that the antisymmetric part of \eqref{6.5} 
\begin{equation}\label{6.6}
- \tilde{\nabla}_h K_{[ji]}^{\;\;\;\;h} + 2k{\tilde\nabla}_h T^{h}_{\;\;[ij]} = \Sigma_{[ij]}
\end{equation}
amounts to the conservation laws for the spin \eqref{5.11}. Indeed, in view of eqs. \eqref{3.1.20} and \eqref{6.3}, eqs. \eqref{5.11} are seen to reduce to
\begin{equation}\label{6.7}
\frac{1}{2}\tilde{\nabla}_h S_{ij}^{\;\;\;h} + \Sigma_{[ij]} =0
\end{equation}
Inserting eqs. \eqref{3.1.12} into \eqref{6.7} and taking the identity $T_{ij}^{\;\;\;h}= K_{ji}^{\;\;\;h} - K_{ij}^{\;\;\;h} $ into account, we end up with eqs. \eqref{6.6}.

Saturating eqs. \eqref{6.7} with $U_i$ we obtain the explicit expression for the vector density of energy-momentum
\begin{equation}\label{6.8}
P^j = \rho U^j + S^{ij}U^h\tilde{\nabla}_h U_i
\end{equation}
where $\rho := U^iP_i$. Inserting eqs. \eqref{6.8} into \eqref{6.1a}, we get the form of the energy-momentum tensor
\begin{equation}\label{6.9}
\Sigma^{ij} = \left(\rho + p\right) U^iU^j - p g^{ij} + U^iS^{pj}U^h\tilde{\nabla}_h U_p
\end{equation}
The significant part of the Einstein-like equations turns out to be the symmetric one
\begin{equation}\label{6.10}
\begin{split}
\tilde{R}_{ij} -\frac{1}{2}\tilde{R}g_{ij} = \Sigma_{(ij)} - \frac{(1-6k)}{4(1-4k) (1+2k)}S_{pq}S^{pq}U_iU_j +\\
+ \frac{(1-6k)}{8(1-4k) (1+2k)}S_{pq}S^{pq}g_{ij} - \frac{1}{2(1+2k)}\tilde{\nabla}_h\left(S^h_{\;\,i}U_j + S^h_{\;\,j}U_i\right)
\end{split}
\end{equation}
\subsubsection{FLRW cosmological models}
In the above described theory, Friedmann-Lema\^{\i}tre-Robertson-Walker (FLRW) cosmological models 
\begin{equation}\label{6.4.8.0}
ds^2 = dt^2 - \frac{a^{2}(t)}{\left(1 + Kr^2/4\right)^2} \left(dr^2 + r^2\,d\theta^2 + r^2\sin^{2}\theta\,d\varphi^2 \right)
\end{equation}
(with $K=-1,0,1 $) can be considered under the hypothesis that the isotropic and homogeneous universe is filled with an unpolarized spinning cosmological fluid. Being the spin randomly oriented, we can assume that the average of the spin and its gradient vanish, but the same is not true for the spin-squared terms $<S^{pq}S_{pq}>:=\frac{1}{2}S^2 $. The conclusion follows that, after averaging, eqs. \eqref{6.10} reduce to 
\begin{equation}\label{6.4.8.1}
\begin{split}
\tilde{R}_{ij} -\frac{1}{2}\tilde{R}g_{ij} = \Sigma_{(ij)} - \frac{(1-6k)}{4(1-4k) (1+2k)}S_{pq}S^{pq}U_iU_j +\\
+ \frac{(1-6k)}{8(1-4k) (1+2k)}S_{pq}S^{pq}g_{ij} 
\end{split}
\end{equation}
and the energy-momentum tensor is 
\begin{equation}\label{6.4.8.2}
\Sigma_{ij} = (\rho + p)\,U_iU_j - p\,g_{ij}
\end{equation}
The Friedmann equations derived from \eqref{6.4.8.0}, \eqref{6.4.8.1} and \eqref{6.4.8.2} are
\begin{subequations}\label{6.4.8.3}
\begin{equation}\label{6.4.8.3a}
\ddot{a} = \frac{a}{6}\left[ -\rho -3p + 4C(k)S^2 \right]
\end{equation}
\begin{equation}\label{6.4.8.3b}
\dot{a}^2 = \frac{a^2}{3}\left[ -\frac{3K}{a^2} +\rho - C(k)S^2 \right]
\end{equation}
\end{subequations}
where we have defined $C(k):=\frac{(1-6k)}{4(1-4k)(1+2k)}\/$. Under the stated conditions, it is easily seen that the conservation laws \eqref{5.15} and \eqref{5.11} give rise respectively to the equations
\begin{subequations}\label{6.4.8.3.1}
\begin{equation}\label{6.4.8.3.1a}
\dot\rho + 3\left(\rho + p\right)\frac{\dot a}{a}=0
\end{equation}
\begin{equation}\label{6.4.8.3.1b}
\dot{S} + 3S\frac{\dot a}{a}=0
\end{equation}
\end{subequations}
Supposing again a state equation of the kind $p=\lambda\rho$ ($0\leq\lambda<1$), eqs. \eqref{6.4.8.3.1} admit general solutions
\begin{equation}\label{6.4.8.3.2}
\rho = \rho_0 a^{-3(1+\lambda)}, \qquad S=\frac{S_0}{a^3}
\end{equation}
$\rho_0$ and $S_0$ being suitable integration constants. Eqs. \eqref{6.4.8.3.1} ensure that the quadridivergence (with respect to the Levi--Civita covariant derivative) of the Einstein-like equations \eqref{6.4.8.1} vanishes, while eqs. \eqref{6.4.8.3.2} yield the relation 
\begin{equation}\label{6.4.8.3.2.1}
S^2 = S_0^2\/\left(\frac{\rho}{\rho_0}\right)^{\frac{2}{1+\lambda}}
\end{equation}
We can see from the previous equations that the spin contributions introduce two important modifications in cosmology. The first one is that torsion may open closed universes, 
as a matter of fact from equation (\ref{6.4.8.3b}) it follows that the critical density is defined by
\begin{equation}
\label{criticaldensity }
\rho_{c}=H^{2}+C(k)S^2
\end{equation} 
accordingly,  those  cosmological models with density $\rho$ such that 
\begin{equation}
\label{ inbetween}
H^{2}<\rho<H^{2}+C(k)S^2
\end{equation}
are closed in general relativity, but open in our theory.

The second important modification is that from eq. \eqref{6.4.8.3b} is easily seen that the spin contributions can avoid the initial singularity. Indeed, supposing for simplicity the metric \eqref{6.4.8.0} spatially flat ($K=0$) as well as $C(k)>0$, from eqs. \eqref{6.4.8.3b}, \eqref{6.4.8.3.2} and \eqref{6.4.8.3.2.1} we derive the existence of a minimum value $a_i$ for the scale factor 
\begin{equation}\label{6.4.8.4}
a_i = \left[\frac{C(k)S_0^2}{\rho_0}\right]^{\frac{1}{3(1-\lambda)}}
\end{equation}
corresponding to a maximum value $\rho_i$ of the energy density
\begin{equation}\label{6.4.8.5}
\rho_i = \left[ \frac{\rho_0^{\frac{2}{1+\lambda}}}{C(k)S_0^2} \right]^{\frac{1+\lambda}{1-\lambda}}
\end{equation}
Moreover, from eq. \eqref{6.4.8.3a} it is seen that we can have an initial accelerated expansion ($\ddot a >0$) for values of energy density
\begin{equation}\label{6.4.8.6}
\rho > \rho_f = \left[\frac{(1+3\lambda)}{4C(k)S_0^2}\rho_0^{\frac{2}{1+\lambda}}\right]^{\frac{1+\lambda}{1-\lambda}}
\end{equation}
The value of the scale factor $a_f$ associated with $\rho_f$ is
\begin{equation}\label{6.4.8.7}
a_f = \left[\frac{4C(k)S_0^2}{(1+3\lambda)\rho_0}\right]^{\frac{1}{3(1-\lambda)}}
\end{equation}
Since $\rho_i > \rho_f$, the accelerated expansion stops when density $\rho_f$ and factor scale $a_f$ are reached. It is worth noticing that, while the values of $a_i$ and $a_f$ (as well as of $\rho_i$ and $\rho_f$) depend on the parameter $k$, their ratio $a_f/a_i$ does not and it is identical to the result already obtained in the standard Einstein--Cartan framework \cite{Gasperini}.

In view of this, the cases $K\not = 0$ seem then more interesting. To discuss this point, we limit to consider a fluid with an equation of state of radiation kind $\displaystyle{\lambda}=1/3$. Accordingly, from eq. \eqref{6.4.8.3b} we find the minimal value $a_i$ by solving the equation ($\dot a =0$)
\begin{equation}
\label{2nddegreeequation}
C(k)S_{0}^{2}\frac{1}{a^{4}}-\rho_{0}\frac{1}{a^{2}}+3K=0
\end{equation}
The solutions are
\begin{equation}
\label{seconddegreesolution}
 \frac{1}{a^{2}}=\frac{\rho_{0}\pm \sqrt{\rho_{0}^{2}-12KC(k)S^{2}_{0}}}{2C(k)S^{2}_{0}}\ \ \mathrm{or} \ \ \ \ a=\sqrt{\frac{2C(k)S^{2}_{0}}{\rho_{0}\pm \sqrt{\rho_{0}^{2}-12KC(k)S^{2}_{0}}}}
\end{equation} 
When $K=1$, these solutions make sense only if 
\begin{equation}
\label{ validityC(k)}
C(k)\le \frac{\rho_{0}^{2}}{12S^{2}_{0}}
\end{equation}
representing a further restriction on the parameter $k$ (together with the requirement $C(k)>0$). The two solutions \eqref{seconddegreesolution} coincide with the minimum and the maximum reachable values of the scale factor of the universe. 

For $K=-1$ the imaginary solution 
\begin{equation}
\label{kless0 }
 a=\sqrt{\frac{2C(k)S^{2}_{0}}{\rho_{0}-\sqrt{\rho_{0}^{2}+12C(k)S^{2}_{0}}}}
\end{equation}
has no physical meaning. So there is only a minimum
\begin{equation}\label{min}
a_i = \sqrt{\frac{2C(k)S^{2}_{0}}{\rho_{0} + \sqrt{\rho_{0}^{2}+12C(k)S^{2}_{0}}}}
\end{equation}
Anyway, in both cases $K=\pm 1$, from eq. \eqref{6.4.8.3a} it is seen that the universe can undergo an accelerated expansion for $a_i \leq a \leq a_f$, where
\begin{equation}\label{af}
a_f = \sqrt{\frac{2C(k)S_0^2}{\rho_0}}
\end{equation}
In the case of a closed universe the final expansion is limited by the condition (\ref{ validityC(k)}). In the case of an open universe the ratio
\begin{equation}\label{ratio}
\frac{a_f}{a_i} = \sqrt{\frac{\rho_0 + \sqrt{\rho_0^2 + 12C(k)S_0^2}}{\rho_0}}
\end{equation}
depends actually on the parameter $k$. Note that $C(k)\rightarrow +\infty$ for $k\rightarrow -\frac{1}{2}^{+}$. We conclude that in this case the parameter $k$ may be chosen in such a way that the accelerated expansion is sufficiently long to predict a flat observed universe.
 
At the  end of this section with add some  final remarks about the horizon problem, one of the classical problems of cosmology. 

It is well-known that in the general relativistic cosmological models the horizon problem is a consequence of the initial singularity.  In fact, in general relativity, from the Friedmann equation
\begin{equation}
\label{Friedmann equation}
\dot{a}^2 = \frac{a^2}{3}\left[ -\frac{3K}{a^2} +\rho \right]
\end{equation}
it follows that $\dot{a}\to \infty$ when $a\to0$. This velocity must be compared with 
\begin{equation}
\label{null equation}
\frac{dr}{dt}=\pm \frac{c}{a}
\end{equation}
which has been obtained by imposing the condition $ds^{2}=0$; this means that the signals exchanged in a fluid with equation of state $p=\lambda\rho$ by different parts of the universe cannot go faster than $c/a$ , while the various regions of the universe move away more rapidly, when $a$ is very small. It follows that the particle horizon
\begin{equation}
\label{hor}
l(t_0)= a(t_{0})\int_{0}^{t_{0}}\frac{c dt}{a(t)} 
\end{equation}
is finite, while $H\sim1/t$ is not an integrable function in the same interval. 
The consequence is  that there are regions of the universe  which are not connected causally. In General Relativity this problem has been solved, together with the flatness problem and other classical problems by introducing the inflationary scenario.

But comparing equations  (\ref{Friedmann equation}) and (\ref{null equation}) we see that  as the universe expands, there is an epoch when
\begin{equation}
\label{inversion }
\frac{c}{a}> v=\frac{\dot{a}}{a}d
\end{equation}
where $v$ is the recession velocity  related to the distance  $d$  by the Hubble law.
So assuming in our theory  a similar expansion law,  but without  the initial singularity and  assuming that  $C(k)$ is  large enough, we can show that  horizon defined  by equation (\ref{hor}) can cover a region large enough such that all the observed parts of the universe are in mutual causal connection. This explains why all the regions of the universe show the same temperature and the same physical properties.
\section{Conclusion}
In this paper, we have constructed a generalization of the Einstein-Cartan gravity in which both curvature and torsion are considered each with its own coupling constant, the one related to the curvature being the gravitational constant while the one related to torsion still undetermined; we have applied this geometrical background to the case of Dirac fields and spin fluids, showing that the corresponding equations have torsional contributions whose coupling constant can be suitably chosen: in the case of the coupling to Dirac spinors, we have shown, through two examples, that the torsional contributions with a properly tuned coupling constant make the non-linearities of the Dirac equation relevant already at subatomic scales; in the case of the spin fluid, where in cosmological models the torsional contributions were already present, the tuning of the coupling constant amplifies them even more: in particular, in the case of spatially hyperbolic FLRW model, the accelerated expansion lasts longer, solving the horizon problem in a natural way.

A problem this approach may not be able to address is the fact that there does not seem to be a unique value of the constant for which all of the applications we mentioned fit into the presented frame, and so only an even more general theory of gravity may be able to provide a running coupling constant, possibly scaling with the energy, such that all applications above, and maybe more, can fit into a single scheme \cite{FV}.

If this approach of ours, as well as any of its generalizations, really works, then torsion would not only be observable, but it might have already been observed, although we have not been able to recognize it for what it is.

We are aware of the fact that torsion may not be the answer to all, and maybe not even to most, of the problems of physics, but it may possibly be for some of them, and our theory can allow us to see how.

\end{document}